# Block Sparse Memory Improved Proportionate Affine Projection Sign Algorithm

J. Liu, and S. L. Grant

A block sparse memory improved proportionate affine projection sign algorithm (BS-MIP-APSA) is proposed for block sparse system identification under impulsive noise. The new BS-MIP-APSA not only inherits the performance improvement for block-sparse system identification, but also achieves robustness to impulsive noise and the efficiency of the memory improved proportionate affine projection sign algorithm (MIP-APSA). Simulations indicate that it can provide both faster convergence rate and better tracking ability under impulsive interference for block sparse system identification as compared to APSA and MIP-APSA.

*Introduction:* Adaptive filters have been widely used in various applications of system identification in which the normalized least mean square (NLMS) algorithm is well-known due to its simplicity, but suffers from slow convergence for colored input [1]. The affine projection algorithm (APA) provides better convergence for colored input compared with NLMS [2]. Meanwhile, the family of affine projection sign algorithm (APSA) has been proposed to improve the performance of APA under impulsive noise together with lower complexity [3]. In order to exploit the sparsity of some echo paths, the real-coefficient improved proportionate APSA (RIP-APSA) was proposed [4], and a memory improved proportionate APSA (MIP-APSA) was further proposed to achieve improved steady-state misalignment with similar computational complexity compared with RIP-APSA [5]. Recently, the block-sparse improved proportionate NLMS (BS-IPNLMS) algorithm was proposed to improve the performance of IPNLMS for identifying block-sparse systems [7]. In this Letter, motived by both BS-PNLMS and MIP-APSA, we will propose a block sparse memory improved proportionate APSA (BS-MIP-APSA) algorithm, which not only inherits the performance improvement for block-sparse system identification, but also achieves robustness to impulsive noise and the efficiency of MIP-APSA.

*Review of MIP-APSA:* For echo cancellation, the far-end signal $\mathbf{x}(n)$ is filtered through the echo path $\mathbf{h}(n)$ to get the desired signal $y(n)$.

$$y(n) = \mathbf{x}^T(n)\mathbf{h}(n) + v(n) \qquad (1)$$

$$\mathbf{x}(n) = [x(n)\, x(n-1) \cdots x(n-L+1)]^T, \qquad (2)$$

$$\mathbf{h}(n) = [h_0(n)\, h_1(n) \cdots h_{L-1}(n)]^T, \qquad (3)$$

super-script $T$ denotes transposition, $L$ is the filter length, $n$ is the time index, and $v(n)$ is the background noise plus near-end signals. Let $\hat{\mathbf{h}}(n)$ be the $L \times 1$ adaptive filter coefficient vector which estimates the true echo path vector $\mathbf{h}(n)$ at iteration $n$, and group the $M$ most recent input vectors together:

$$\mathbf{X}(n) = [\mathbf{x}(n)\, \mathbf{x}(n-1) \cdots \mathbf{x}(n-M+1)], \qquad (4)$$

$$\mathbf{e}(n) = \mathbf{y}(n) - \mathbf{X}^T(n)\hat{\mathbf{h}}(n-1), \qquad (5)$$

$$\mathbf{y}(n) = [y(n)\, y(n-1) \cdots y(n-M+1)]^T, \qquad (6)$$

where $M$ is called the projection order. In [5], MIP-APSA proposed the following weight update:

$$\mathbf{g}(n) = [g_0(n), g_1(n), \cdots, g_{L-1}(n)], \qquad (7)$$

$$g_l(n) = \frac{(1-\alpha)}{2L} + \frac{(1+\alpha)|\hat{h}_l(n)|}{2\sum_{i=0}^{L-1}|\hat{h}_i(n)| + \varepsilon}. \qquad (8)$$

$$\mathbf{P}(n) = [\mathbf{g}(n) \odot \mathbf{x}(n), \mathbf{P}_{-1}(n-1)], \qquad (9)$$

$$\mathbf{x}_{gs}(n) = \mathbf{P}(n)\mathrm{sgn}(\mathbf{e}(n)), \qquad (10)$$

$$\hat{\mathbf{h}}(n+1) = \hat{\mathbf{h}}(n) + \frac{\mu \mathbf{x}_{gs}(n)}{\sqrt{\delta + \mathbf{x}_{gs}^T(n)\mathbf{x}_{gs}(n)}}, \qquad (11)$$

where $-1 \leq \alpha < 1$, $l = 0,1,\cdots,L-1$, $\varepsilon$ is a small positive constant that avoids division by zero, the operation $\odot$ denotes the Hadamard product, $\mathbf{P}_{-1}(n-1)$ contains the first $M-1$ columns of $\mathbf{P}(n-1)$, $\mathrm{sgn}(\cdot)$ takes the sign of each element of a vector, and $\varepsilon$ is a small positive constant. Compared with RIP-PAPSA, MIP-PAPSA takes into account the 'proportionate history' from the last $M$ moments of time. More details can be found in [5]-[6].

*Algorithm design:* In network echo cancellation, the network echo path is typically characterized by a bulk delay dependent on network loading, encoding, and jitter buffer delays and an "active" dispersive region in the range of 8-12 *ms* duration [1]. Meanwhile, it is well-known that NLMS is preferred over PNLMS for dispersive system. Therefore, considering the block-sparse characteristic of the network impulse response, the BS-PNLMS algorithm was proposed to improve the PNLMS algorithm by exploiting this special block-sparse characteristic, in which BS-PNLMS used the same step-size within each block and the step-sizes for each block were proportionate to their relative magnitude [7]. We propose to take in account the block-sparse characteristic and partition the MIP-APSA adaptive filter coefficients into $N$ groups with group-length $P$, and $L = N \times P$,

$$\hat{\mathbf{h}}(n) = [\hat{\mathbf{h}}_0(n), \hat{\mathbf{h}}_1(n), \cdots, \hat{\mathbf{h}}_{N-1}(n)], \qquad (12)$$

then the control matrix $\mathbf{g}(n)$ in (7)-(8) is be replaced by

$$\tilde{\mathbf{g}}(n) = [\tilde{g}_0(n)\mathbf{1}_P, \tilde{g}_1(n)\mathbf{1}_P, \cdots, \tilde{g}_{N-1}(n)\mathbf{1}_P] \quad (13)$$

$$\tilde{g}_k(n) = \frac{(1-\alpha)}{2L} + \frac{(1+\alpha)\|\hat{\mathbf{h}}_k(n)\|_2}{2N\sum_{i=0}^{N-1}\|\hat{\mathbf{h}}_i(n)\|_2 + \varepsilon}, \quad (14)$$

in which $\mathbf{1}_P$ is a $P$-length column vector of all ones, and $\|\hat{\mathbf{h}}_k(n)\|_2 = \sqrt{\sum_{j=1}^{P} \hat{h}_{kN+j}^2(n)}$, $k = 0,1,\cdots,N-1$. The weight update equation for BS-MIP-ASPA is

$$\tilde{\mathbf{P}}(n) = [\tilde{\mathbf{g}}(n) \odot \mathbf{x}(n), \tilde{\mathbf{P}}_{-1}(n-1)], \qquad (15)$$

$$\tilde{\mathbf{x}}_{gs}(n) = \tilde{\mathbf{P}}(n)\mathrm{sgn}(\mathbf{e}(n)), \qquad (16)$$

$$\hat{\mathbf{h}}(n) = \hat{\mathbf{h}}(n-1) + \frac{\mu \tilde{\mathbf{x}}_{gs}(n)}{\sqrt{\delta + \tilde{\mathbf{x}}_{gs}^T(n)\tilde{\mathbf{x}}_{gs}(n)}} \qquad (17)$$

where $\tilde{\mathbf{P}}_{-1}(n-1)$ also contains the first $M-1$ columns of $\tilde{\mathbf{P}}(n-1)$. It should be noted that the proposed BS-MIP-APSA includes both APSA and MIP-APSA. The MIP-APSA algorithm is a special case of proposed BS-MIP-APSA with group length $P = 1$. Meanwhile, when $P$ is chosen as $L$, the BS-MIP-APSA algorithm degenerates to APSA.

*Complexity:* Compared with traditional RIP-APSA and MIP-APSA, the extra computational complexity of the BS-MIP-APSA arises from the computation of the $l_2$ norm in (14), which requires $L$ multiplications and $N$ square roots. The complexity of the square root could be reduced through a look up table or Taylor series [7]. Meanwhile, the increase in complexity can be offset by the performance improvement as shown in the simulation results.

*Simulation results:* In our simulation, the echo path is a $L = 512$ finite impulse response (FIR) filter, and the adaptive filter is the same length. We generated colored input signals by filtering white Gaussian noise through a first order system with a pole at 0.8. Independent white Gaussian noise is added to the system background with a signal-to-noise ratio (SNR) of 40 dB. The impulsive noise with signal-to-interference ratio (SIR) of 0 dB is generated as a Bernoulli-Gaussian (BG) distribution. BG is a product of a Bernoulli process and a Gaussian process, and the probability for Bernoulli process is 0.1. The performance was evaluated through the normalized misalignment: $10\log_{10}\left(\|\mathbf{h}-\hat{\mathbf{h}}\|_2^2 / \|\mathbf{h}\|_2^2\right)$. In order to evaluate the tracking ability, we switch the echo path from the one-cluster block-sparse system of Fig. 1(a) to the two-cluster block-sparse system of Fig. 1(b).

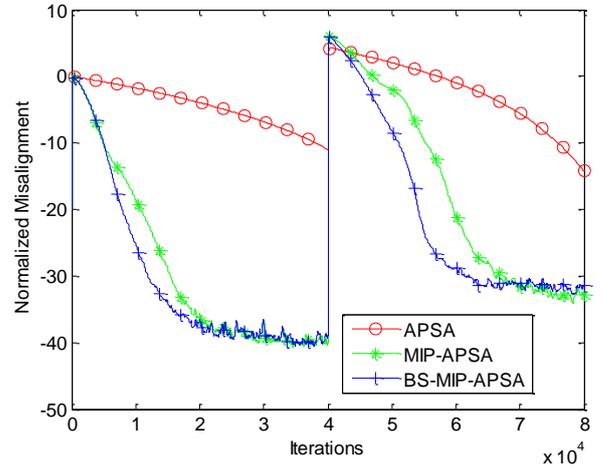

**Fig. 2** *Normalized misalignment of APSA, MIP-APSA, and BS-MIP-APSA for colored input signal.*

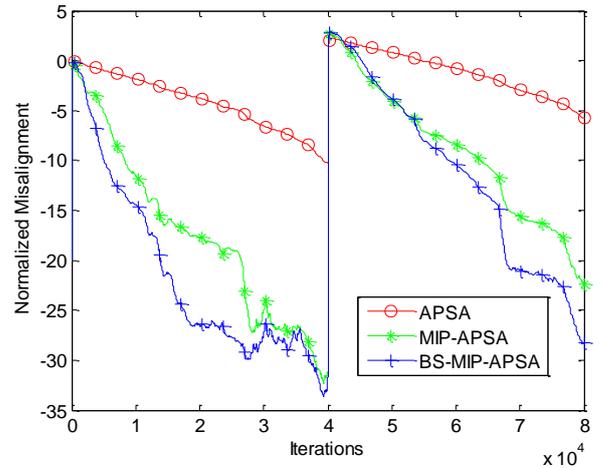

**Fig. 3** *Normalized misalignment of APSA, MIP-APSA, and BS-MIP-APSA for speech input signal.*

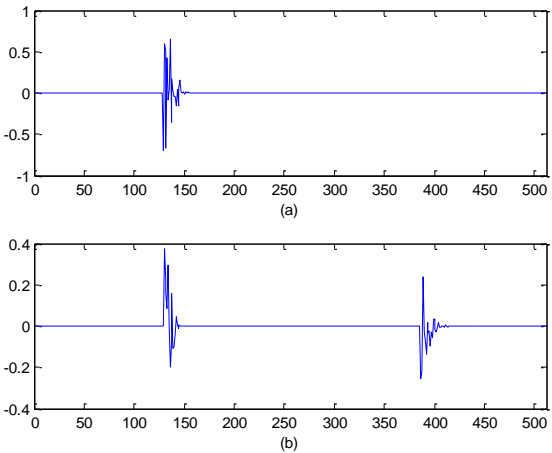

**Fig. 1** *Two block-sparse systems used in the simulations: (a) one-cluster block-sparse system, (b) two-cluster block-sparse system.*

The APSA and MIP-APSA algorithms are compared with BS-MIP-APSA. The parameters are $\mu = 0.001$, $\varepsilon = 0.01$, $\delta = 0.01$, $\alpha = 0$, $M = 2$, and $P = 4$. In the first case, we show the normalized misalignment for colored input in Fig. 2. We could see that the proposed BS-MIP-APSA achieves both faster convergence rate and better tracking ability. In Fig. 3, the performance of BS-MIP-APSA is compared with APSA and MIP-APSA for speech input signal, and we found that our proposed algorithm demonstrates better performance too.

*Conclusion:* We have proposed a block-sparse memory improved affine projection sign algorithm to improve the performance of block-sparse system identification. Simulations demonstrate the proposed algorithm has both faster convergence speed and tracking ability for block-sparse system identification compared with APSA and MIP-APSA algorithms.

*Acknowledgments:* This work is supported by Wilkens Missouri Endowment.

J. Liu, and S. L. Grant (*Department of Electrical and Computer Engineering, Missouri University of Science and Technology, Rolla, MO, USA*)
E-mail: sgrant@mst.edu